\documentclass[showpacs,prb,manuscript,aps]{revtex4}
\draft
\usepackage{graphicx}

\begin{document}
\title{Spin-phonon coupling and pressure effect in the superconductor LiFeAs :
Lattice dynamics from first-principles calculations}
\author{G. Q. Huang$^{1}$, Z. W. Xing$^{2}$, D. Y. Xing$^{2}$}
\address{$^1$Department of Physics and Institute of Theoretical Physics,
Nanjing Normal University, Nanjing 210046, China\\
$^2$National Laboratory of Solid State Microstructures, Nanjing
University, Nanjing 210093,
 China}

\begin{abstract}
The lattice dynamics and the effect of pressure on superconducting
LiFeAs in both nonmagnetic (NM) and striped antiferromagnetic (SAF)
phases are investigated using the plane-wave pseudopotential,
density-functional-based method. While the obtained electron-phonon
coupling $\lambda$  is very small for the NM calculation, the
softening of phonon in the SAF phase may lead to a large increase in
$\lambda$. In the SAF phase, strong anisotropy of the phonon
softening in the Fe plane is found to arise from different spin
orders in the $x$ and $y$ directions, indicating that the phonon
softening is of spin-phonon coupling origin. For the SAF structure,
the calculated variation trend of the electronic density of states
and the phonon frequencies under pressure can explain a large
negative pressure coefficient of $T_{c}$ in the LiFeAs compound.

\end{abstract}



\maketitle
\section{introduction}
\par
The discovery of superconductivity in  layered iron
pnictides\cite{1} has received tremendous attention. The
superconducting transition temperature in doped \textit{Re}FeAsO
systems (\textit{Re}=Ce, Pr. Nd, Sm, ...) was quickly raised above
50 K.\cite{2,3} Other series including the doped
\textit{A}Fe$_2$As$_2$ (\textit{A} = alkali-earth),~\cite{4,5} doped
SrFeAsF,~\cite{6,7,8} doped FeSe,~\cite{9,10,11,12} and LiFeAs
\cite{13,14,15,16}compounds were also found to be superconductors.
All these compounds have formal FeAs layer, which consists of a
square Fe lattice with As in the center of the square but being
alternately shifted above and below the Fe plane.  Both experimental
and theoretical results showed that FeAs layer is the conducting
layer, and dominates the main features near the Fermi
level.\cite{17,18,19}

\par
The superconducting mechanism in iron pnictides is  still under
debate. Since the electron-phonon (EP) interaction in the
paramagnetic phase could only account for a maximum $T_{c}$ of 0.8
K,~\cite{18}  theoretical calculations~\cite{18,20} ruled out phonon
mediated superconductivity. On the other hand, McGuire \textsl{et
al.}~\cite{21} suggested that strong electron-phonon coupling
existed in the high temperature tetragonal phase of LaFeAsO, as
evidenced by the behavior of the mobility, thermal conductivity, and
Seebeck coefficient through the tetragonal-orthorhombic
phase-transition region. The isotope effect~\cite{22} was large in
the pnictide superconductors, implying that the EP interaction
should play an important role in the superconducting mechanism.
Egamin \textsl{et al.}~\cite{23} suggested that the EP coupling
through the spin channel might be sufficiently strong to be an
important part of the superconductivity mechanism in Fe pnictides.

\par
The undoped material ReFeAsO,~\cite{1,24,25}
BaFe$_2$As$_2$,~\cite{26} SrFe$_2$As$_2$,~\cite{27} and
SrFeAsF,~\cite{28} have been reported to undergo a spin-density wave
transition. Upon doping,  the spin-density wave is suppressed and
superconductivity emerges.~\cite{1,24,25,26,27,28} First-principles
calculations of LaFeAsO,~\cite{29} BaFe$_2$As$_2$,~\cite{30} and
SrFeAsF~\cite{31} all showed that the striped-antiferromagnetic
(SAF) configuration is the stable ground state in these compounds.
It was reported that, unlike the known other undoped intrinsic FeAs
compounds, LiFeAs did not show any spin-density wave behavior but
exhibits superconductivity at ambient pressures without chemical
doping.~\cite{13,14,15,16} But very recently, the experimental
results provided direct evidence of the magnetic ordering in the
nearly stoichiometric NaFeAs.~\cite{32} The calculations for the
nonmagnetic (NM) phase showed that the Fermi surface and band
structures near Fermi level of LiFeAs are very similar to other iron
pnictides.~\cite{30,33} And the first-principles
calculations~\cite{34,35} showed that stoichiometric LiFeAs has
almost the same SAF spin order as other FeAs-based parent compounds.

\par
Many high pressure experiments on iron pnictides have been reported
in the literature. The superconductivity can be induced by pressure
in undoped iron pnictides.~\cite{36,37} For the doped iron
pnictides, which are superconducting at ambient pressure, the
pressure can change their critical transition
temperatures.~\cite{25,38} For LiFeAs, the experimental measurements
showed that the superconducting $T_{c}$ decreases linearly with
pressure at a rate of $-1.5$ K/GPa.~\cite{39,40,41}

\par
In this paper we first calculate the electronic structure, phonon
spectrum and the electron-phonon interaction for the NM phase of
LiFeAs compounds. The results obtained are consistent with those in
previous works.~\cite{30,33} Since it is expected that the
superconductivity is related to the spin order in the system,  we
investigate the lattice dynamics in the case of the SAF spin order.
Our results show that the phonons from vibrations of the Fe and As
atoms are softened due to the spin-lattice interaction, so as to
enhance the electron-phonon coupling in the iron pnictides. The
pressure effects for both NM and SAF phases are also investigated.

\section{ Computational Method }
\par
The calculations have been performed in a plane-wave
pseudo-potential representation through the PWSCF program.~\cite{42}
The ultrasoft pseudo-potential~\cite{43}  and general gradient
approximation (GGA-PBE)~\cite{44}  for the exchange and correlation
energy functional are used with a cutoff of 30 Ry for the wave
functions and 240 Ry for the charge densities. For the electronic
structure calculations, the Brillouin zone  integrations are
performed by using the Gaussian smearing technique with a width of
0.04 Ry. Within the framework of the linear response theory, the
dynamical matrixes and  the electron-phonon interaction coefficients
are calculated. The energy and frequency convergence are checked
with respect to the cutoff energy and  k-point sampling.

\section{ Results and discussion}
\subsection{Calculation in the NM phase}
We first do the first-principles calculations for LiFeAs in the NM
case, and then compare our results with the previous
works.~\cite{30,33} The LiFeAs compound  crystallizes in a
PbFCl-type structure (space group $P4/nmm$). The Wyckoff positions
for Fe, Li, and As are $2b$, $2c$, and $2c$, respectively.
Structural optimizations involving the lattice constants and the
internal coordinates are performed using the
Broyden-Fletcher-Goldfarb-Shanno algorithm.~\cite{45} The obtained
lattice constants, $(a, c) = (3.83, 6.55){\AA}$, are slightly larger
than the experimental values $(3.79, 6.36){\AA}$.~\cite{13} The
obtained internal coordinate, $z_{Li}$ = 0.3464, is close to the
experimental value of 0.3459,~\cite{13} but the obtained
$z_{As}$=0.2083 is noticeably lower than the reported value of
$z_{As}$=0.2365.~\cite{13} The resulting height difference is about
0.15 ${\AA}$.

\par
The electronic structure calculation agrees well with previous
calculations for the non-spin-polarized situation. LiFeAs belongs to
be the low carrier density metal with high density of states. The
calculated  density of states at the Fermi level, $N(E_{F})$, is
equal to 3.71 eV$^{-1}$ per unit cell (two formula units), which is
in between 3.86 eV$^{-1}$ obtained in Ref. \onlinecite{33} and 3.58
eV$^{-1}$ in Ref. \onlinecite{30}. The states near the Fermi level
are dominated by Fe $3d$ states lightly mixed with As $p$ states.
The Fermi surface consists of hole cylinders centered at the zone
center and electron cylinders centered at the zone corner. In
general, the electronic structure near the Fermi level of LiFeAs is
qualitative similar to that of the other FeAs materials.

\par
The calculated phonon dispersion of LiFeAs along major high symmetry
directions of the Brillouin zone is plotted in  Fig. 1. The
vibrations of Fe and As atoms occupy the whole energy range, while
the vibrations of Li atoms occupy the high-frequency region because
of its very small mass. We notice that the Fe-As bond-stretching
modes are on the side of high frequency. In fact, the modes having
the same symmetry may be coupled with each other. The phonon
eigenvectors have a strongly mixed character and cannot be simply
traced back to a single vibration pattern. In the high-frequency
region, our calculated phonon spectrum is different from that
obtained previously by Jishi \textsl{et al.}.~\cite{46}  The top
bands for the vibrations of Li atoms are separated from other bands
in Ref. \onlinecite{46}, while they are coupled with other bands in
our calculations.  This difference may stem from different lattice
constants and atomic internal coordinates used in calculations. In
this paper, we use the optimized lattice constants and atomic
internal coordinates, while Jishi \textsl{et al.}  used the
corresponding experimental values. The frequencies involving the
vibrations of Li atoms are sensitive to computational details due to
very small atomic mass.

\par
 The EP coupling constant $\lambda$ and the
logarithmically averaged frequency $\omega_{ln}$ are directly
obtained by evaluating
\begin{equation}
 \lambda=2\int_{0}^{\infty}\frac{\alpha^{2}F(\omega)}{\omega}d\omega
 ,
\end{equation}
and
\begin{equation}
\omega_{ln}=\exp
\left(\frac{2}{\lambda}\int_{0}^{\infty}\frac{\alpha^{2}F(\omega)ln\omega}{\omega}d\omega\right),
\end{equation}
where the EP spectral function $\alpha^{2}F(\omega)$ can be
determined self-consistently by the linear response theory. The
calculated results are $\lambda$=0.26 and $\omega_{ln}$=236 K. Using
the McMillan expression for $T_{c}$ and taking the Coulomb
pseudopotential parameter $\mu^{*}$= 0.1, the resulting value of
$T_{c}$ is much less than 1 K. This is consistent with the result
reported by Jishi \textsl{et al.}.\cite{46} It then followed that
the electron-phonon coupling is too weak to account for
superconductivity in LiFeAs. However, the phonon mediated
superconductivity cannot be ruled out only according to the
conclusion in the NM case.

\subsection{Spin-lattice interaction }

\par
Recently, Li \textsl{et al.}~\cite{34} investigated all the possible
magnetic orders for stoichiometric LiFeAs by using the
full-potential linearized augmented plane wave method, and concluded
that the magnetic ground state of LiFeAs is an SAF order in each Fe
layer and a weak antiferromagnetic  order in the $z$ direction. In
this paper we have done the same calculations by using the
plane-wave pseudo-potential method. Our results confirm the
conclusion in Ref. \onlinecite{34} that stoichiometric LiFeAs has
almost the same SAF spin order as other FeAs-based parent compounds.
Next, we investigate the lattice dynamics for this SAF spin order.
Since the antiferromagnetic interaction in the $z$ direction is very
weak,~\cite{34} we will not consider it in the following
calculations.

\par
To accommodate the SAF magnetic structure, we use a
$\sqrt{2}\times\sqrt{2}\times1$ supercell. The magnetic structure in
the Fe layer is that the Fe spins align ferromagnetically in the $x$
direction and antiferromagnetically in the $y$ direction. We have
performed the full structural optimization, including the shape and
size of the unit cell as well as the atomic internal positions. The
result is that the lattice structure is distorted from the
tetragonal (higher) symmetry to orthogonal (lower) symmetry with
lattice constants as $(a, b, c) =(5.462, 5.559, 6.776){\AA}$. They
are larger than the experimental values, especially along the $z$
direction; whereas in Refs.\ [34] and [35] the calculated
equilibrium volume of the SAF magnetic structure by the projector
augmented wave method is somewhat smaller than the experimental
values. The obtained internal parameters of Li and As atoms are
0.3504 and 0.2206, respectively. Our calculated heights of the As
atoms from the Fe plane for the NM and SAF structures are 1.364
${\AA}$ and 1.495 ${\AA}$, respectively. The corresponding
experimental value is 1.504 ${\AA}$,~\cite{13} which is closer to
the calculated value in the presence of the SAF ordering. This
conclusion is consistent with those in previous
references~\cite{34,35,47,48,49}. The obtained magnetic moment per
Fe atom by the GGA calculation in this paper is about 2.60 $\mu B$.
According to other studies\cite{34,35}, the  magnetic moment
calculated by the local spin-density approximation is much smaller
than that by the GGA calculation. Next, we calculate all zone-center
phonon frequencies for the SAF magnetic structure within the
framework of the linear response theory. For comparison, we also
calculate the zone-center phonon frequencies for the NM structure
with the same $\sqrt{2}\times\sqrt{2}\times1$ supercell. However, it
is difficult to directly compare the changes of phonon frequencies
 because of different irreducible representations in the two
structures.

\par
To investigate how the  spin order in SAF magnetic structure changes
the phonon frequency in LiFeAs, we define a  weighted average
frequency as
\begin{equation}
\overline{\omega}_{\alpha,n}=\sum_{\nu}\frac{u_{\alpha,n,\nu}^{2}}
{|\bf{u}_{\nu}|^{2}}\omega_{\nu},
\end{equation}
where $\alpha$ is the polarization index ($x,y,z$), $n$ is the
number of atom index, and the sum runs over all the modes.
$|\bf{u}_{\nu}|$ is the amplitude of the atomic displacements for
the $\nu$th normal mode and
${u_{\alpha,n,\nu}^{2}}/{|\bf{u}_{\nu}|^{2}}$ represents the
percentage of the contribution from the $\alpha$ polarization
direction of the $n$th atom. The calculated average frequencies for
the NM and SAF structures at the theoretical zero pressure (P = 0)
are listed in Table 1, in which the relative changes of frequency
($\frac{\overline{\omega}_{SAF}-\overline{\omega}_{NM}}{\overline{\omega}_{NM}}
\times 100\%$) are also given. An evident feature in Table 1 is the
softening of the phonon for Fe and As atoms in the SAF magnetic
structure relative to that in the NM structure. Furthermore, the
softening of phonons is nearly isotropic in the in-plane directions
for the As atoms, but strongly anisotropic for the Fe atom. The most
softening appears for $\overline{\omega}_{x}$ of the Fe atom. It
changes from 216.92 $cm^{-1}$ in the NM structure to 148.68
$cm^{-1}$ in the SAF magnetic structure, reducing about 31\%. By
comparison, $\overline{\omega}_{y}$ of the Fe atom reduces only
about 14\%.

\par
To further investigate behavior of the phonon softening, we have
examined the As $A_{g}$ mode and  Fe $E_{g}$  mode (named as in the
tetragonal structure) with the frozen-phonon calculation. Both the
two modes are coupled with other modes according to the symmetry. In
the frozen-phonon calculation, we don't consider the coupling with
vibrations of other atoms. For the As $A_{g}$ mode, As atoms above
and below the Fe plane displace along $z$ in opposite directions.
The calculated distorted energies versus atom displacements are
plotted in Fig.\ 2a, which can be well fitted to $E = k_2 u^{2}$ of
harmonic modes. It can be clearly seen that the energy curve,
$E=27.70 u^2$, obtained in the NM structure is steeper than $E=13.87
u^2$ in the SAF structure.   The calculated frequencies $\omega$ are
28.29meV and 22.62meV for the NM structure and SAF structures,
respectively, yielding a reducing of about 20\%. Since the
Fe-magnetic moment is sensitive to the As-z position, the calculated
Fe-magnetic moment versus the As displacement is also plotted in the
insert of Fig.\ 2a. With increasing the distance between the As atom
and the Fe plane, the obtained Fe-magnetic moment increases, which
has the same trend in LaOFeAs as calculated by Yildirim~\cite{50}.

\par
For the in-plane  Fe $E_{g}$ mode, from the higher tetragonal
symmetry to lower orthogonal symmetry, the doubly degenerate mode
will split into non-degenerate ones, as shown in the insert of Fig.\
2b. For the doubly degenerate mode in the NM structure, the
distorted energy can be well fitted to quadratic equation  $E=30.04
u^2$, and the obtained frequency is 33.29 meV. For the two
non-degenerate modes in the SAF structure, the distorted energy may
be fitted to $E = k_2 u^{2} + k_4 u^{4}$, where $k_4u^{4}$ is the
anharmonic term of the mode. Using the Hartree-Fock decoupling, one
gets $E(u)= (k_2 + 3 k_4\langle u^{2}\rangle)u^{2}$, where $\langle
u^{2}\rangle =\hbar/2M\omega_{sch}$ and $\omega_{sch}=[(k_2+
3k_4\langle u^{2}\rangle )/M]^{1/2}$ is the self-consistent phonon
frequency.\cite{51} For the Fe atom vibration along the $x$ ($y$)
direction, we obtain $E = 13.70 u^{2} + 59.52 u^{4}$ ($E = 21.71
u^{2} + 25.24 u^{4}$), the harmonic frequency $\omega =22.48$meV
(28.30meV), and the self-consistent phonon frequency $\omega_{sch}=$
22.80meV (28.47meV). The anharmonic effect leads to a slight
enhancement in the phonon frequency. From the NM structure to the
SAF structure, the phonon frequency decreases about 31.5\% (14.5\%)
for the Fe atom vibration along the $x$ ($y$) direction. It is
interesting to note that the present softening of phonon frequencies
obtained by the frozen-phonon calculation is quite well consistent
with that of $\overline{\omega}_{x}$ and $\overline{\omega}_{y}$
obtained above by the linear response perturbation theory.

\par
The conclusion of the phonon softening  in the SAF magnetic phase
agrees well with the results studied by others. The first-principles
calculations by Fukuda \textsl{et al.}~\cite{52} for the NM
structure in LaFeAsO$_{1-x}$F$_{x}$ and PrFeAsO$_{1-y}$ indicated
that the calculated phonon DOS could agree with the experimental
spectrum only when the computed Fe-As force constant is reduced by
30\%. The calculations by Yildirim\cite{50} showed that the in-plane
Fe-Fe and $c$-polarized As phonon modes are softened about 10\% for
the LaOFeAs system, and softened about 10-14\% and 23\% for the
BaFe$_{2}$As$_{2}$,  explaining the experimental data~\cite{53,54}.
As a result, it can be concluded that the SAF magnetic structure has
significant effects on the phonon frequency, which may be a common
feature in iron pnictides.

\par
The difference of lattice constants $a$ and $b$ in the orthogonal
structure is very small ($b/a=1.02$), but the difference of the
phonon softening along $x$ and $y$ direction for Fe atoms is large.
This seems to indicate that the difference in phonon softening
arises from the anisotropic spin order rather than the change of the
geometric structure. In order to confirm this point, we have carried
out calculations by the linear response perturbation theory for both
NM and magnetic SAF phases, in which the lattice parameters are
fixed to the experimental values of the tetragonal structure and
other internal parameters are optimized. The calculated results for
the ambient pressure are listed in the top rows of Table 1. It is
found that there is large phonon softening for the vibrations of
both Fe and As atoms; and that for the Fe atoms, the magnitude of
softening in the $x$ direction is larger than that in the $y$
direction.   It then follows that the spin-lattice interaction is
the origin of the softening. It is the different spin orders in the
$x$ and $y$ directions that result in strong anisotropy of the
phonon softening in the Fe plane.

\par
Owing to the presence of strong spin-lattice interactions, it is
imperative to study the phonon-mediated mechanism via the spin
channel. Yildirim~\cite{29} suggested that magnetism and
superconductivity in doped LaFeAsO may be strongly coupled, much
like in the high-Tc cuprates. The large iron isotope effect found by
Liu \textsl{et al.}~\cite{22} for both spin-density wave and
superconducting transition temperatures seemed to indicate that
phonons not only play a dominant role for superconductivity, but
also are close coupled with magnetism in the iron pnictides. Egami
\textsl{et al.}~\cite{23} suggested that the EP coupling through the
spin channel may be sufficiently strong to be an important part of
the superconductivity mechanism in Fe pnictides. According to Eq.\
(1), we can see that the softening of phonons in the SAF phases is
favorable to increasing the EP coupling constant $\lambda$, and an
increasing $\lambda$ will lead to an enhancement of $T_{c}$.

\subsection{Pressure effect}
For LiFeAs, the experimental measurements showed that the
superconducting $T_{c}$ decreases linearly with pressure at a rate
of $-1.5$ K/GPa.~\cite{39,40,41} Here we investigate the pressure
effect for both NM and SAF phases. Constant pressure molecular
dynamics is performed to optimize the cell size, volume, and atomic
internal positions. The optimized parameters together with some
calculated results are  listed in Table 2. The lattice constants,
the volume of the unit supercell, and the Fe-As bond length decrease
monotonically with increasing the pressure up to 15 GPa. The
decrease of $c/a$ under pressure indicates that the out-of-plane
compression is larger than the in-plane ones. From Table 2, we can
see that the decrease of percentage of the volume in the SAF phase
is larger than that in the NM phase under the same pressure,
indicating that the compression in the SAF phase is easier than that
in the NM phase.  For example, from the theoretical zero pressure to
1.5Gpa, the volume decreases 4.88 \% in the SAF structure while it
decreases only 3.31 \% in the NM structure. We also calculate the
bulk modulus $B$ by fitting the $E$ (total energy) versus $V$
(volume) curves to the Birch-Murnagham equation of state. The
obtained $B$ are 454 and 353 kbar for the NM and SAF, respectively.
The smaller bulk modulus $B$ in the SAF structure means weaker
interactions between atoms and smaller force constants,  resulting
in a lower frequency of phonon. This is just consistent with the
calculated results discussed above.

\par
The density of states at the Fermi level, $N(E_{F})$,  is also
listed in Table 2. In the range up to 15Gpa,  $N(E_{F})$  decreases
with compression in the SAF structure, as expected as usual. In
general, a compression of the lattice by pressure causes an increase
of the bandwidth, which in turn results in a decrease of the
averaged density of states. For the NM phase, the Fermi level lies
just in the flat region of the DOS curve, and so $N(E_{F})$ changes
little under pressure.  For the SAF structure, the calculated
magnetic moment each iron atom decreases as the pressure is
increased, which has the same trend as the calculation by  Nakamura
\textsl{et al.}~\cite{54} for the LaFeAsO compound.

\par
Next, we wish to study the pressure effect on the phonon frequency.
The calculated average frequencies at p= 1.5 Gpa and 15 Gpa for the
two structures are also listed in Table 1. As usual, all the
calculated frequencies increase with pressure, for  a smaller
lattice constant leads to a stronger force constant. For the
calculated average frequencies of the Fe and As atoms, however, the
increase in the SAF structure is much larger than that in the NM
structure. From Table 1, one also finds that Up to 15Gpa, the phonon
softening from the NM to SAF phase and its anisotropy in the Fe
plane still exist, but such a phonon softening decreases with the
increase of pressure. Relative to those in the NM structure, the
calculated average frequencies of the Fe atom in the SAF structure
reduce about 31.5\% and 13.8\% for $\overline{\omega}_{x}$ and
$\overline{\omega}_{y}$, respectively, at the theoretical zero
pressure. Correspondingly, they reduce only about 12.5\% and 6.4\%
for $\overline{\omega}_{x}$ and $\overline{\omega}_{y}$,
respectively, at 15Gpa. It then follows that the difference in some
physical properties between the NM and SAF structures becomes small
with increasing pressure. The SAF structure may not be the ground
structure when the pressure is increased to some extent, as has been
found in LaOFeAs~\cite{54}.

\par
For the SAF structure, our calculations show that with increasing
pressure, $N(E_{F})$ decreases and the phonon frequencies increase
greatly, resulting in a large decrease of the EP coupling constant
$\lambda$. All these factors will lead to a lower $T_{c}$ and a
large negative pressure coefficient of $T_{c}$. As a result, the
pressure effect on superconducting $T_{C}$ is not contradictory to
the EP coupling mechanism provided that the spin degree of freedom
is taken into account.

\subsection{Summary} The electronic structure, phonon spectrum and
the EP interaction in the NM LiFeAs compound have been investigated
by the first-principles calculations. The obtained electron-phonon
coupling for the NM calculation is too weak to account for
superconductivity in LiFeAs. After considering the spin order, the
calculated average frequencies, especially from the vibrations of Fe
and As atoms in the SAF structure, are smaller than those in  the NM
structure. It is found that the different spin orders in the $x$ and
$y$ directions result in strong anisotropy of the softening of
phonons in the Fe plane. It then follows that the origin of the
phonon softening is the spin-lattice interaction. The softening of
phonons in the SAF phase can lead to an increase of EP coupling
constant $\lambda$ and  superconducting temperature $T_{c}$. By the
calculations under different pressures, the obtained bulk modulus
$B$ in the SAF structure is smaller than that in the NM structure,
indicating that the compression in the SAF phase is easier than that
in the NM phase. For the SAF structure, our calculations show that
with increasing pressure, both the decrease of the density of states
at the Fermi level and the increase of  the phonon frequencies make
$\lambda$ and $T_{c}$ decrease, resulting in a large negative
pressure coefficient of $T_{c}$.

\section*{Acknowledgments}

G.Q.H. and D.Y.X. acknowledge support from the National Natural
Science Foundation of China under Grant No. 10947005, Z.W.X
acknowledges support from the Natural Science Foundation of Jiangsu
Province in China under Grant No. SBK200920627, and G.Q.H.
acknowledges support from the Foundation of Jiangsu Education Office
of China under Grant No. 09KJB140004. This work is also supported by
the State Key Program for Basic Researches of China under Grants No.
2006CB921803 and No. 2010CB923400.

\newpage
\begin{table}
\caption{\label{tab:table1} The calculated average frequency
(cm$^{-1}$) at different pressures. For the ambient pressure, the
lattice parameters are fixed to the experimental values of the
tetragonal structure and the internal parameters are optimized. }

\begin{ruledtabular}
\begin{tabular}{cccc}
&$(\overline{\omega}_{x},\overline{\omega}_{y},\overline{\omega}_{z})$
for
Fe&$(\overline{\omega}_{x},\overline{\omega}_{y},\overline{\omega}_{z})$
for
As&$(\overline{\omega}_{x},\overline{\omega}_{y},\overline{\omega}_{z})$
for Li\\
ambient pressure&&&\\
NM phase &(224.69, 224.69, 207.50)& (173.70, 173.70,
183.27)&(233.02,
233.02, 299.80)\\
SAF phase&(175.44, 212.30, 192.59) &(152.15, 149.95, 164.09)& (232.65, 245.07, 330.25)\\
$\frac{\overline{\omega}_{SAF}-\overline{\omega}_{NM}}{\overline{\omega}_{NM}}\times100$(\%)&( $-$21.92, $-$5.51, $-$7.19)&($-$12.41, $-$13.67, $-$10.47)&($-$0.l6, 5.17, 10.16)\\
\\
P = 0 (Gpa)&&&\\
NM phase &(216.92, 216.92, 198.59)&(170.36, 170.36, 174.69)&(210.01, 210.01, 249.66)\\
SAF phase&(148.68, 187.03, 170.22)&(141.73, 139.17, 142.20)&(194.88, 191.49, 221.98)\\
$\frac{\overline{\omega}_{SAF}-\overline{\omega}_{NM}}{\overline{\omega}_{NM}}\times100$(\%)&($-$31.46, $-$13.78, $-$14.29)&($-$16.81, $-$18.31, $-$18.60)&($-$7.20, $-$8.82, $-$11.09)\\
\\
P = 1.5 (Gpa)&&&\\
NM phase &(220.47, 220.47, 203.67)&(172.64, 172.64, 180.29)&(225.82 225.82, 282.33)\\
SAF phase&(158.20, 192.67, 176.43)&(145.51, 142.48, 151.46)&(213.77, 205.11, 267.77)\\
$\frac{\overline{\omega}_{SAF}-\overline{\omega}_{NM}}{\overline{\omega}_{NM}}\times100$(\%)&($-$28.24, $-$12.61, $-$13.37)&($-$15.71, $-$17.47, $-$15.99)&($-$5.34, $-$9.17, $-$5.16)\\
\\
P = 15 (Gpa)&&&\\
NM phase &(249.15, 249.15, 232.44)&(186.26, 186.26, 208.41)&(312.79, 312.79, 427.45)\\
SAF phase&(218.02, 233.29, 218.76)&(166.53, 160.47, 186.53)&(328.32, 300.24, 413.71)\\
$\frac{\overline{\omega}_{SAF}-\overline{\omega}_{NM}}{\overline{\omega}_{NM}}\times100$(\%)&($-$12.49, $-$6.37, $-$5.89)&($-$10.59, $-$13.85, $-$10.50)&(4.96, $-$4.01, $-$3.21)\\

\end{tabular}
\end{ruledtabular}
\end{table}

\newpage
\begin{table}
\caption{\label{tab:table2}Crystal structure data and the density of
states at the Fermi level $N(E_{F})$ with
$\sqrt{2}\times\sqrt{2}\times1$ supercell under different pressures.
 $V_{0}$ means the theoretical equilibrium volume }

\begin{ruledtabular}
\begin{tabular}{ccccccc}
p(Gpa)&(Exp)\footnotemark[1]&0&0.5&1.0&1.5&15\\
NM phase &&&&&&\\
a(${\AA}$)&5.3610&5.4168&5.4078&5.3950&5.3795&5.1645\\
c/a&1.1872&1.2094&1.2042&1.2011&1.1939&1.1578\\
V(${\AA}^{3}$)&182.920&192.220&190.441&188.605&185.863&159.484\\
(V-V$_{0}$)/V$_{0}\times100$(\%)&&0&$-$0.93&$-$1.88&$-$3.31& $-$17.03\\
z(As)&0.2365&0.2083&0.2096&0.2110&0.2136&0.2355\\
z(Li)&0.3459&0.3465&0.3466&0.3469&0.3481&0.3445\\
$d_{Fe-As}(\AA)$&2.4204&2.3516&2.3491&2.3467&2.3451&2.3058\\
$N(E_{F}$)(States/eV)&&7.424&7.372&7.346&7.350&7.456\\
\\
SAF phase&&&&&&\\
a(${\AA}$)&&5.4616&5.4194&5.4039&5.3890&5.0712\\
b/a&&1.0178&1.0183&1.0186&1.0187&1.0205\\
c/a&&1.2405&1.2357&1.2340&1.2273&1.2142\\
V(${\AA}^{3}$)&&205.693&200.282&198.353&195.669&161.598\\
(V-V$_{0}$)/V$_{0}\times100$(\%)&&0&$-$2.631&$-$3.568&$-$4.873& $-$21.437\\
z(As)&&0.2206&0.2242&0.2247&0.2263&0.2439\\
z(Li)&&0.3504&0.3532&0.3541&0.3550&0.3496\\
$d_{Fe-As}(\AA)$&&2.4554&2.4481&2.4421&2.4369&2.3530\\
$N(E_{F}$)(States/eV)&&3.446&3.350&3.302&3.116&2.802\\
\end{tabular}
\end{ruledtabular}

\footnotetext[1]{from Ref.~\onlinecite{13}.}
\end{table}

\begin{figure}
\includegraphics{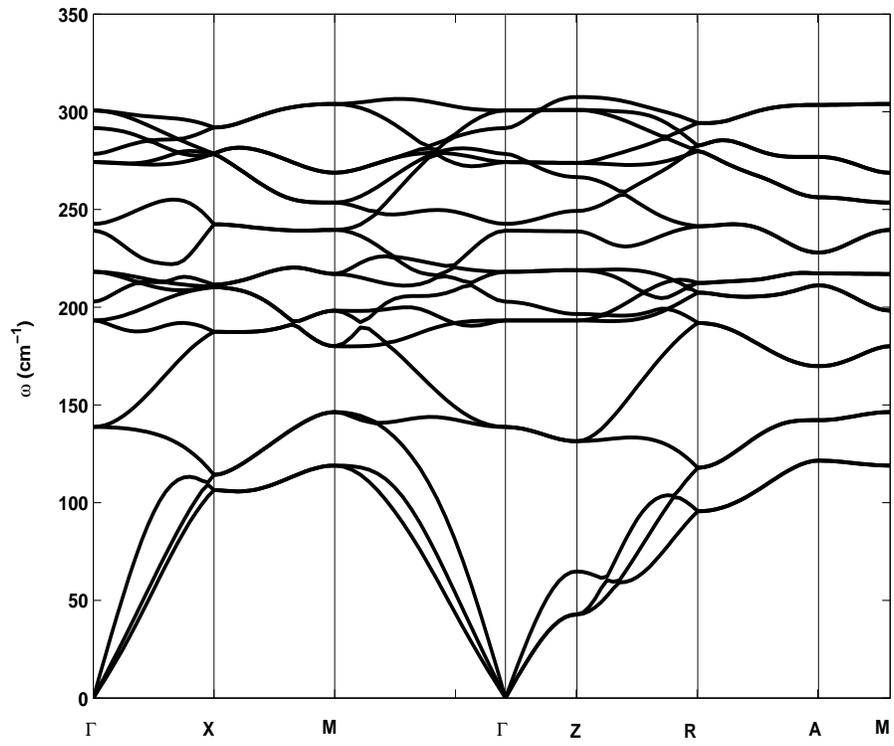}
 \caption{\label{fig:epsart}
The calculated phonon dispersion curves for LiFeAs }
\end{figure}

\begin{figure}
\includegraphics{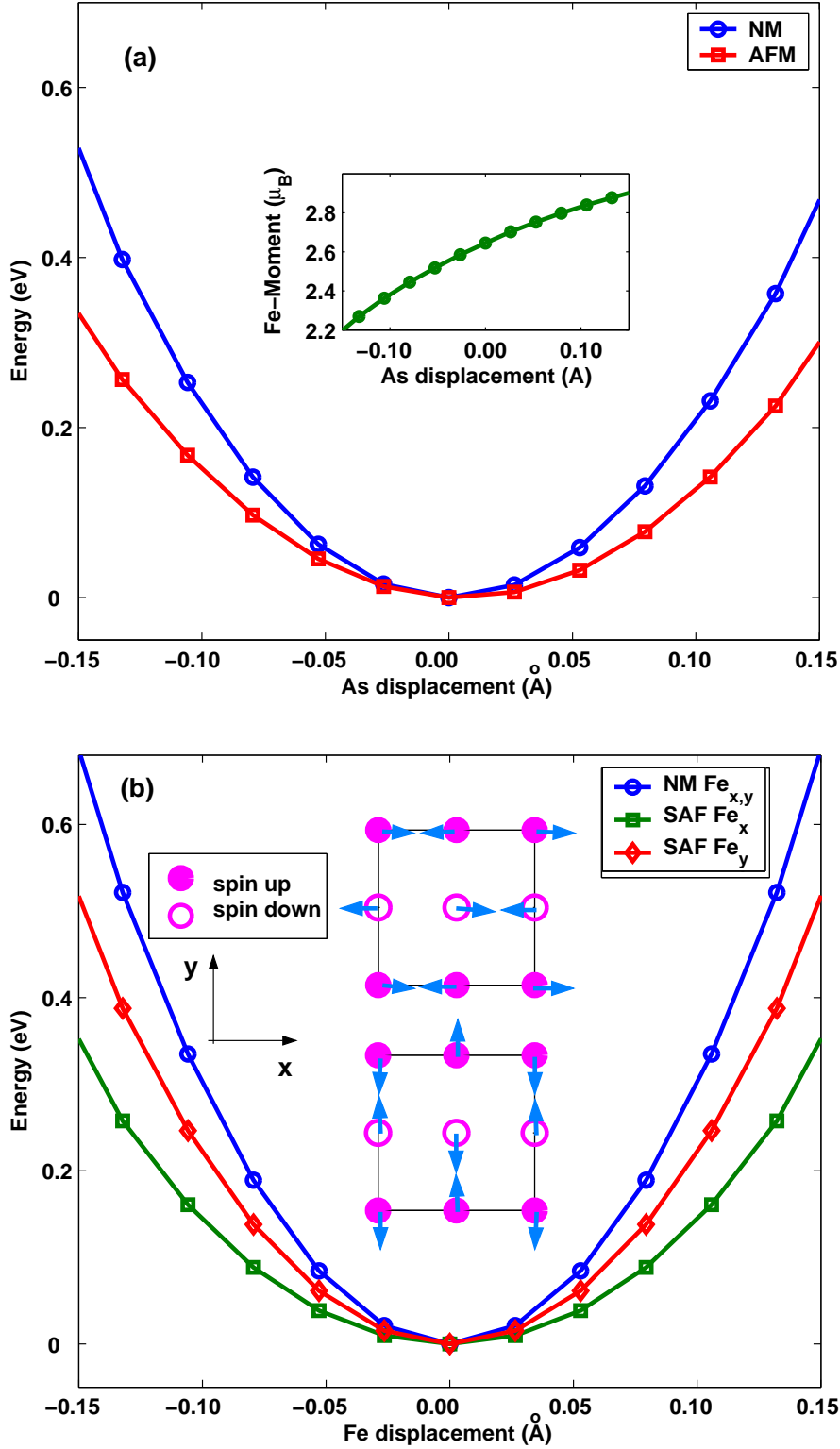}
 \caption{\label{fig:epsart}
Energy curves as a function of atom displacement. (a) for  As
$A_{g}$ mode and (b) for Fe $E_{g}$  mode.  The calculated
Fe-magnetic moment vs. As displacement is shown in  the insert of
(a), Negative As displacement corresponds to As atoms moving towards
the Fe-plane. Phonon displacement patterns for the in-plane  Fe
$E_{g}$ mode are shown in the insert of (b).}
\end{figure}

\end{document}